\documentclass{article}
\usepackage{amsmath}
\usepackage{bm}
\usepackage[left=1in,top=1in,right=1in,bottom=1in]{geometry}
\usepackage{subcaption,graphicx}
\usepackage{titlesec}
\usepackage{comment}
\usepackage[section]{placeins}
\usepackage{amsfonts}
\usepackage{algorithm}
\usepackage{algpseudocode}
\usepackage{tikz}
\usepackage{csquotes}
\usepackage{caption}
\usepackage{appendix}
\MakeOuterQuote{"}
\usepackage{todonotes}
\usepackage{ulem}
\usepackage{setspace}
\title{\textbf{Joint Reconstruction of Multi-channel, Spectral CT Data via Constrained Total Nuclear Variation Minimization}}
\author{ David S. Rigie and Patrick J. La Rivi\`{e}re}

\DeclareMathOperator*{\argmin}{argmin}

\begin{document}
\bibliographystyle{ieeetr}

\newcommand{\utrue}{\bm{u}_\mathrm{true}}
\newcommand{\epsref}{\epsilon_{\mathrm{ref}}}
\newcommand{\Grad}[1]{\mathrm{Grad}\,#1}
\newcommand{\Div}[1]{\mathrm{Div}\,#1}
\newcommand{\TV}{\mathrm{TV}}
\newcommand{\VTV}{\mathrm{VTV}}
\newcommand{\TNV}{\mathrm{TNV}}
\newcommand{\inner}[1]{\left<#1\right>}
\renewcommand{\div}[1]{\mathrm{div}\, #1}
\newcommand{\unoisy}{\bm{u}_{\mathrm{noisy}}}
\newcommand{\uu}{{\bm{u}}}
\newcommand{\uubar}{{\bar{\bm{u}}}}
\newcommand{\zz}{{\bm{z}}}
\newcommand{\qq}{{\bm{q}}}
\newcommand{\deltak}{\delta_{\mathcal{N}}}
\newcommand{\jth}{j^{\mathrm{th}}}
\newcommand{\ff}{\bm{f}}
\newcommand{\xx}{\bm{x}}
\newcommand{\yy}{\bm{y}}
\newcommand{\prox}[2]{\mathbf{prox}_{#1}\hspace{-2pt}\left(#2\right)}
\renewcommand{\vec}[1]{\mathrm{vec}\,#1}

\maketitle

\begin{abstract}
\textit{
We explore the use of the recently proposed "total nuclear variation" (TNV) \cite{Rigie2014,Holt2014} as a regularizer for reconstructing multi-channel, spectral CT images. This convex penalty is a natural extension of the total variation (TV) to vector-valued images and has the advantage of encouraging common edge locations and a shared gradient direction among image channels. We show how it can be incorporated into a general, data-constrained reconstruction framework and derive update equations based on the first-order, primal-dual algorithm of Chambolle and Pock. Early simulation studies based on the numerical XCAT phantom indicate that the inter-channel coupling introduced by the TNV leads to better preservation of image features at high levels of regularization, compared to independent, channel-by-channel TV reconstructions. 
}
\end{abstract}

\section{Introduction}
Since the influential work of Rudin, Osher, and Fatemi \cite{Rudin1992}, the total-variation (TV) has been used for a variety of image processing tasks, such as denoising \cite{Rudin1992}, deconvolution \cite{Chan1998}, and demosaicking\cite{Komatsu2005}. Because the TV is a nonsmooth penalty that disfavors oscillations while allowing for sharp discontinuities, it has been found to effectively suppress noise while preserving edges. The TV has more recently been rediscovered as a prior for tomographic inverse problems suffering from incomplete projection data \cite{Candes2006,Sidky2006}. Consequently, many works have investigated TV-regularized reconstruction algorithms for sparse-view, limited angle, and low-dose CT \cite{Bian2013,Song2007,Tian2011}. 

In last couple of years, several papers have proposed generalizations of the TV for multi-channel or vector-valued images, in order to extend popular image-processing techniques to the RGB color space \cite{Sapiro1997,Blomgren1998,Goldluecke2012,Lefkimmiatis2013,Holt2014}. Coincidentally, there is also a fast-growing interest in multi-spectral CT imaging, due to rapid developments in energy-resolving, photon-counting detectors. Since spectral CT is essentially just CT with multiple "color" channels, many of these theoretical advances can be directly applied to spectral CT reconstruction.

The primary consideration when formulating a vectorial TV (VTV) is how the color channels should be coupled. One could na\"{\i}vely apply the conventional TV to each channel independently, but this would fail to leverage the strong inter-channel correlations inherent to multi-spectral images. In this work, we advocate for a VTV that penalizes the nuclear norm of the Jacobian derivative because it has several desirable properties. This convex prior encourages all image channels to have common edge locations as well as for their gradient vectors to point in the same direction. The notion that gradient vectors of different color channels should be approximately aligned has been suggested previously \cite{keren1998,Holt2011,Rigie2014}. Additionally, we will demonstrate that the inter-channel coupling helps prevent edge smearing in the noisiest color channels, allowing for stronger regularization without significant feature loss. This specific VTV has been recently proposed by Lefkimmiatis \cite{Lefkimmiatis2013}, Rigie \cite{Rigie2014}, and Holt \cite{Holt2014}. We borrow from Holt in referring to it as the "Total Nuclear Variation" (TNV). 

The main objective of this paper is to demonstrate the usefulness of the TNV as a regularizer for multi-energy CT reconstruction. Our study contains several important contributions. To our knowledge, this is the first study with empirical results demonstrating the use of the TNV for spectral CT reconstruction, and our results indicate a clear advantage over independent, channel-by-channel reconstructions. Secondly, we propose a novel, noise-balancing transform that improves noise-suppression for multi-channel data for which the color channels have mismatched noise levels. This is a situation that is likely to occur in photon-counting CT, where the lowest and highest energy windows tend to suffer from relatively low count rates. Finally, we demonstrate that the TNV can be applied not only to the energy-binned data, but also to synthesized basis-material data. In most color image processing tasks, the measured RGB basis is the same as the observed basis. The analog in spectral CT would be to directly reconstruct the energy-bin data, so we will refer to this as the "color" basis. However, often the goal is, instead, to reconstruct density maps of different basis materials \cite{Fessler2002,Long2014}, which can be obtained through a non-linear transformation of the measured data \cite{Alvarez1976}. In this work, we will refer to this space as the "material" basis. We show empirically that the TNV works well in either basis and that it may actually be advantageous to co-reconstruct the color basis and the material basis. We refer to this as the "hybrid" basis, which is formed by concatenating the color and material basis channels. By simultaneously reconstructing the low-noise, color basis and the high-noise, material basis, the TNV is able to better suppress much of the noise that is amplified during the decomposition. 

First, we will review some theoretical concepts and define the total nuclear variation (TNV). Next, we will describe our data-constrained, TNV reconstruction model and show how the first-order, primal-dual algorithm of Chambolle and Pock \cite{Chambolle2011} can be used to derive efficient update equations. We compare our TNV reconstruction model to performing separate channel-by-channel reconstructions using simulated, dual-energy and photon-counting data using the anthropomorphic XCAT phantom \cite{Segars2010}. In the photon-counting study, we present reconstruction results from the color basis, where the color channels correspond to five equispaced energy windows. In the dual kVp study, we present reconstructions of the hybrid basis, where 80 kVp, 140 kVp, bone, and soft-tissue images are simultaneously reconstructed. Overall, we find that the TNV regularization results in less edge smearing at the same noise levels, while the computation time is very similar to the na\"{\i}ve channel-by-channel reconstruction. Additionally, the hybrid TNV reconstruction technique seems to be a promising method for controlling noise in the basis-material images.

\section{Theory}

\subsection{Notation and definitions}
In this work, we deal only with 2D images, but the generalization to 3D is straightforward. Consider a single-energy CT image vector $u \in\mathbb{R}^{\mathcal{L}}$, where $\mathcal{L}$ is the total number of image pixels.  We will also refer to its discrete derivative $Du\in \mathbb{R}^{2\cdot\mathcal{L}}$, which consists of a 2D derivative vector $\left[Du\right]_\ell$ at every pixel location. where $[.]_\ell$ denotes the a particular pixel-index. This quantity is also directly related to the discrete gradient by the following relationship: 
\begin{equation}
\left[Du \right]_\ell^T = \left[\nabla u \right]_\ell.
\end{equation}
 The quantities $u$ and $Du$ are both 1D vectors, so we will use $\|.\|$ and $\inner{.,.}$ to refer to the usual vector norm and inner product. We will also refer to a multi-channel image, $\bm{u} \in \mathbb{R}^{\mathcal{M}\cdot\mathcal{L}}$, with $\mathcal{M}$ spectral channels and its discrete derivative $D\bm{u} \in \mathbb{R}^{(\mathcal{M}\times 2)\cdot\mathcal{L} }$. The quantity $D\bm{u}$ is a discrete analog to the Jacobian derivative of a vector-function, and at every pixel location the first-order derivatives are fully characterized by the $\mathcal{M}\times 2$ rectangular matrix $\left[D\bm{u} \right]_\ell$. For two tensors $\bm{x},\bm{y} \in \mathbb{R}^{(\mathcal{M}\times 2)\cdot\mathcal{L}}$, we define the inner product $\inner{\xx,\yy} = \inner{\vec{\xx},\vec{\yy}}$, where $\mathrm{vec}(.)$ unfolds a tensor into a 1D vector.

\subsection{The scalar total variation}
Total variation regularized CT reconstruction algorithms have been studied extensively due to the approximate gradient sparsity of CT images. Many works have demonstrated that such schemes can often yield high quality images from severely undersampled projection data \cite{Candes2006,Sidky2006,Han2012,Bian2013,Xu2014}. 

\paragraph{The anisotropic TV}
The "anisotropic" TV is simply defined as the $\ell_1$-norm of the derivative of the image $u$,
\begin{equation}
\TV_a(u) = \|Du\|_1,
\end{equation}
where it is useful to think of the $\|.\|_1$ as a "sparsity-inducing" norm because it is a convex relaxation of $\|.\|_0$.
\paragraph{The isotropic TV}
Somewhat more common is the "isotropic" TV, which is defined in terms of the gradient-magnitude image,
\begin{equation}
\TV(u) = \sum_{\ell=1}^\mathcal{L} \|\left[D u\right]_\ell\|_2
\end{equation}
This penalty function has similar properties to the anisotropic TV, but is rotationally invariant, whereas the anisotropic TV tends to under-penalize horizontal and vertical edges. 

\subsection{The vectorial total variation}
Now, we consider generalizing the TV to the $\mathcal{M}$-channel image $\bm{u}$.
\paragraph{Na\"{\i}ve generalization to multi-channel images}
The simplest way to extend the definition of the TV to this vector-valued image, $\bm{u}$, would be to take the summation of the total-variation of each channel. This is given by
\begin{equation}\label{eq:VTVS}
\TV_S(\bm{u}) = \sum_{m=1}^{\mathcal{M}} \TV(\bm{u}_m).
\end{equation}
Since this penalty is additively separable, this would be akin to just reconstructing each channel separately, so we will refer to this approach as the "channel-by-channel" TV. 

\paragraph{A class of vectorial TV regularizers.}

Now, we will describe a more principled way to generalize the TV to multi-channel images. We mentioned earlier that the first-order derivatives of $\bm{u}$ can be characterized at every pixel-location by the Jacobian matrix $\left[D\bm{u} \right]_{\ell}$. An interesting class of regularizers is described by
\begin{equation}\label{eq:VTV}
\VTV(\bm{u}) = \sum_{\ell=1}^{\mathcal{L}} \|\left[D\bm{u}\right]_\ell\|_{S_p},
\end{equation}
where the matrix norm $\|.\|_{S_p}$ is known as the "Schatten p-norm," which is equivalent to the $\ell_p-$norm of the singular values, $\|X\|_{S_p} = \|\sigma(X)\|_{p}$, where $\sigma(X)$ is a vector comprised of the singular values of matrix $X$. Some special cases include $S_\infty$ (spectral norm), $S_2$ (Frobenius norm), and $S_1$ (nuclear norm). All three of these norms have been suggested as potential VTV penalties \cite{Blomgren1998,Goldluecke2012,Lefkimmiatis2013,Holt2014,Rigie2014}, and they share a few important properties. Because they are norms defined by the singular values of the Jacobian, they are rotationally invariant and convex. Additionally, for a single-channel image, they all reduce exactly to the conventional, isotropic TV. 

\paragraph{The Total Nuclear Variation (TNV)}
The total nuclear variation (TNV), defined as the nuclear norm of the Jacobian, is the only one of these VTV's that encourages the gradient vectors of all of the image channels to point in a common direction. We have found empirically that this is a very useful prior for spectral CT images, and that it generally outperforms the other VTV penalties. Therefore, in this work we will focus solely on the TNV, which is defined by 
\begin{equation}\label{eq:TNV}
\TNV(\bm{u}) = \sum_{\ell=1}^{\mathcal{L}} \|\left[D\bm{u}\right]_\ell\|_{S_1}.
\end{equation}
To see how this TNV couples the different energy channels of $\bm{u}$, note that the Jacobian at pixel $\ell$ can also be expressed in terms of the gradient vectors of the various image channels:
\begin{equation}\label{eq:Jacobian}
\left[D\bm{u}\right]_\ell = \begin{bmatrix}
\longleftarrow & \left[\nabla \uu_1\right]_\ell^T & \longrightarrow \\
\longleftarrow & \vdots & \longrightarrow \\
\longleftarrow & \left[\nabla \uu_{\mathcal{M}}\right]_\ell^T & \longrightarrow
\end{bmatrix}_{\mathcal{M}\times\mathcal{N}}.
\end{equation}
\begin{figure}[htbp]
\centering
\newcommand{\pwidth}{0.24\linewidth}
\subcaptionbox{All singular values are zero}{\includegraphics[width=\pwidth]{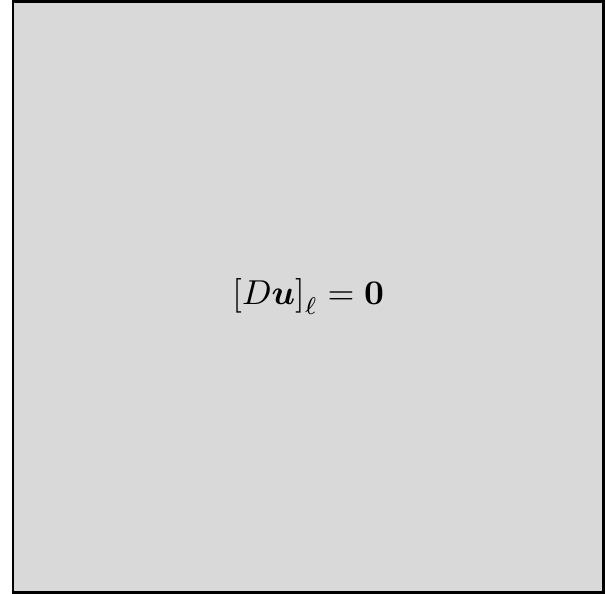}} \hspace{0.5cm}
\subcaptionbox{1 non-zero singular value}{\includegraphics[width=\pwidth]{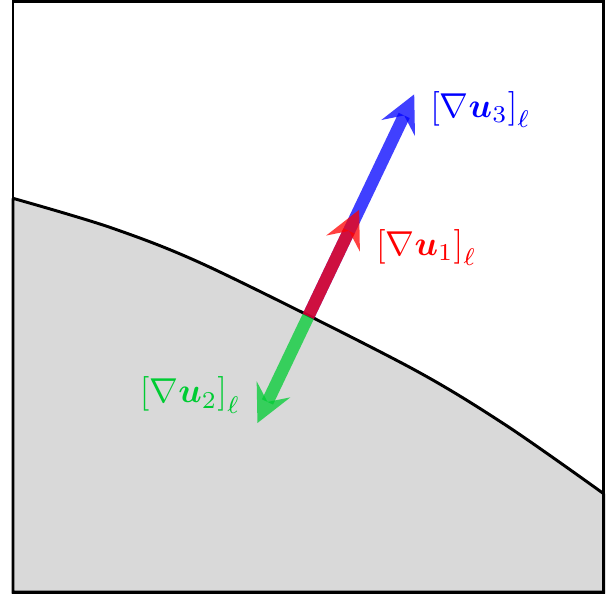}} \hspace{0.5cm}
\subcaptionbox{All non-zero singular values}{\includegraphics[width=\pwidth]{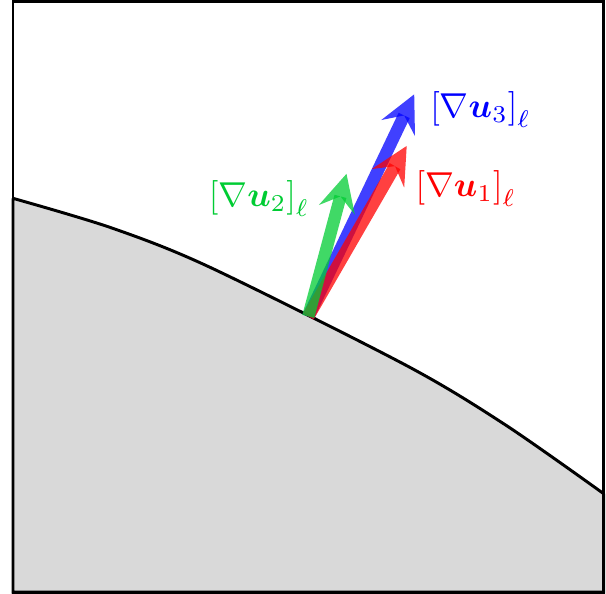}}
\caption{This illustrates the relationship between the directionality of the gradient vectors of a 3-channel image and the singular values of its Jacobian. When all image channels are constant valued, all singular values of the Jacobian are zero (a). If the gradient vectors are parallel or anti-parallel, one singular value is non-zero (b). All singular values will be nonzero if the gradients have unique directions (c).}
\label{fig:gradientdirection}
\end{figure}
\noindent The nuclear norm encourages sparsity in the singular values of $\left[D\bm{u}\right]_\ell$ \cite{Cai2010,Candes2011}. If pixel $\ell$ lies within a constant-valued region of all of the image channels then this Jacobian matrix will be entirely null-valued, and all of its singular values will be zero. Further, if all of the gradient vectors of the various image channels are parallel or anti-parallel, such as at a common edge, then the Jacobian will be rank one and thus have only one non-zero singular value. This is because parallel or anti-parallel vectors are not linearly independent. The fact that the TNV treats parallel and anti-parallel vectors the same means that it is robust to contrast inversions. We also point out that if one of the gradient vectors has zero magnitude, the rank is still one, so the TNV also appears to be robust in handling edges that may not exist in every image channel. In sum, encouraging singular-value sparsity in the Jacobian is akin to encouraging common edge locations and gradient directions. This notion is illustrated graphically in Figure \ref{fig:gradientdirection}.

\paragraph{Converting to a "noise-balanced" image space}

In spectral CT imaging it is often the case that certain color channels are significantly noisier than others, and we have found the TNV is much more effective if a noise-balancing transform is applied prior to reconstruction or denoising. For example, suppose we had a two-channel image, $\bm{u}$, with zero-mean, Gaussian noise with variances given by $\sigma_1^2$ and $\sigma_2^2$ for channels 1 and 2, respectively. If we wanted to use the TNV to denoise $\bm{u}$, we should first define a noise-balanced image $\bm{u'}$ as in equation \ref{eq:noisebalancing}.
\begin{equation}\label{eq:noisebalancing}
\bm{u'} = \begin{bmatrix} u_1/\sigma_1 \\ u_2/\sigma_2 \end{bmatrix}
\end{equation}
Each channel is globally scaled so that the per-channel noise is the same. After denoising, the inverse scale factors are applied in order to return to the original image space. For image reconstruction, the same type of scaling would be applied to the sinogram data. In general, the projection data will not have a uniform noise level within a single color channel, so we use some average measure of channel noise to determine the global scale factor. We find that this noise-balancing procedure is a very important step, as it greatly improves the noise suppression in multi-spectral images with unequal noise levels. In many cases, even images that are corrupted by extremely high noise levels may be well recovered using this technique as long as at least one of the color channels is reasonably clean. 

\begin{figure}[htbp]
\centering
\includegraphics[width=2in]{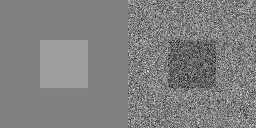}
\includegraphics[width=2in]{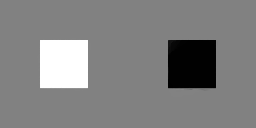}
\caption{A noisy two-channel image (left) with grossly mismatched noise levels. The image is successfully denoised using the TNV in conjunction with the noise-balancing procedure (right). Even the extremely noisy channel is recovered nearly perfectly due to the tight coupling to the clean image channel.}
\end{figure}

\section{Materials and methods}

\subsection{General reconstruction model}

Assume we have a set of $\mathcal{M}$ sinograms, corresponding either to the directly measured energy channels of a spectral CT system (color basis) or to the material basis projections. We jointly reconstruct these channels by solving the following data-constrained optimization problem:
\begin{equation}\label{eq:constrainedVTV}
\begin{split}
\min_{\bm{u}} &\quad \VTV(\bm{u})\\
\mathrm{s.t.} &\quad \|A\bm{u}-\bm{f}\|_W \leq \epsilon,
\end{split}
\end{equation}
where $\ff$ is the multi-channel projection data related to $\uu$ by the discrete fan-beam or cone-beam projection operator $A$. The projector is defined on a per channel basis,
\begin{equation}
\left(A\uu\right)_m = A_m\uu_m,
\end{equation}
where $A_m$ characterizes the measurement geometry associated with channel $m$. The norm on the data constraint $\|.\|_W$ is a weighted $\ell_2$ norm, defined by 
\begin{equation}\label{eq:wl2}
\|x\|^2_W = \inner{x,Wx}.
\end{equation}
This penalized weighted least squares (PWLS) data model is often used in CT reconstruction because when the weights are selected such that $W = K^{-1}_{\bm{f}}$, the data fidelity term is a maximum likelihood estimator for Gaussian data with covariance $K_{\bm{f}}$. The single adjustable parameter $\epsilon$ controls the balance between data-fidelity and regularity. By fixing $\epsilon$ we can directly compare reconstructions using the na\"{i}ve channel-by-channel $\TV_S$ to our proposed $\TNV$ subject to the same data fidelity constraint. 

\subsection{The Chambolle Pock Primal Dual Algorithm}
In this section, we will provide an overview of the first-order, primal-dual algorithm of Chambolle and Pock \cite{Chambolle2011} and demonstrate how to apply it to our reconstruction model. For a tutorial on how to apply the CPPD algorithm to various CT reconstruction schemes, we refer the reader to Sidky et al \cite{Sidky2012}. 

\paragraph{The proximal mapping}
In order to describe the CPPD algorithm, we must first introduce the concept of a proximal mapping. For a function $f(x)$, the proximal mapping is defined by
\begin{equation}\label{eq:prox}
\prox{f}{x} = \argmin_u \,\, f(u) + \frac{1}{2}\|u-x\|_2^2.
\end{equation}
It will also be useful to consider the proximal mapping of the scaled function $\lambda f$, which can be written as
\begin{equation}\label{eq:proxscaled}
\prox{\lambda f}{x} = \argmin_u \,\, f(u) + \frac{1}{2\lambda}\|u-x\|_2^2.
\end{equation}
The proximal mapping can be interpreted as a generalized projection operator because for the special case that $f(x)$ is an indicator function, then it is precisely a euclidean projection. For an extensive overview of the prox operator and applications, refer to Parikh and Boyd \cite{Parikh2013}. 

\paragraph{A general saddle-point problem}

Let $X$ and $Y$ be two finite-dimensional, real vector spaces. The CPPD algorithm is  designed to solve the saddle-point problem described by
\begin{equation}\label{eq:saddle_xy}
\min_{x\in X}\,\max_{y\in Y} \,\, \inner{Kx,y} - F(y) + G(x),
\end{equation}
where $F(y)$ and $G(x)$ are convex functions with very simple proximal mappings, and $K$ is a general linear operator. In particular, the proximal mappings associated with $F$ and $G$ should have a closed form or be easily solvable to high precision. It turns out that many interesting image processing and reconstruction problems fit this description, including the VTV reconstruction model already described in the previous section. The CPPD update equations are summarized by Algorithm \ref{alg:CPPD}.
\begin{algorithm}
\caption{CPPD Algorithm}\label{alg:CPPD}
\begin{algorithmic}[1]
\State Initialize: $\theta\in \left[-1,1\right]$, $\,\, \sigma\tau\|K\|_2^2 < 1$, $\,\, x^{(0)},\bar{x}^{(0)} \in X, \, y^{(0)} \in Y$ \Comment{$\|K\|_2 \equiv \sigma_\mathrm{max}(K)$}
\Repeat
\State $y^{(k+1)} = \prox{\sigma F}{y^{(k)} + \sigma K \bar{x}^{(k)}}$
\State $x^{(k+1)} = \prox{\tau G}{x^{(k)} - \tau K^Ty^{(k+1)}}$
\State $\bar{x}^{(k+1)} = x^{(k+1)} + \theta\left(x^{(k+1)}-x^{(k)} \right)$
\Until $x^{(k+1)},y^{(k+1)} = x^{(k)},y^{(k)}$ \Comment{stop when convergence criteria met}
\end{algorithmic}
\end{algorithm}
The parameters $\theta,\sigma,\tau$ can be thought of as step-size parameters that affect convergence speed but not the final solution, and $\|K\|_2$ is the "spectral norm" of the operator $K$, which is equivalent to its largest singular value. The power method, outlined in Sidky et al \cite{Sidky2012}, provides a very efficient iterative procedure for computing $\|K\|_2$ for very large-scale problems, involving only matrix-vector multiplications. This algorithm has close ties with several other popular algorithms, including ADMM, split-bregman, and proximal-point. This is further discussed in Chambolle and Pock \cite{Chambolle2011}.

\paragraph{Applying the CPPD algorithm to VTV reconstruction}

Now we will outline how the CPPD algorithm can be applied to the data-constrained optimization problem of (\ref{eq:constrainedVTV}). First we rewrite (\ref{eq:constrainedVTV}) as 
\begin{equation}
\min_{\uu} \quad \VTV(\uu) + \delta_{\epsilon}\left(A\uu-\ff\right),
\end{equation}
where the indicator function $\delta_\epsilon(\bm{x})$ is defined by
\begin{equation}\label{eq:dataconstraint}
\delta_\epsilon(\xx) = \left\{\begin{array}{lr} 0 & \|\xx\|_W \leq \epsilon\\
\infty &  \|\xx\|_W > \epsilon
\end{array}\right..
\end{equation}
Now, we introduce an auxiliary variable $\qq$ and rewrite the data constraint as an optimization problem:
\begin{equation}\label{eq:dataconjugate}
\delta_\epsilon(A\uu-\ff) = \max_\qq \,\, \inner{\qq,A\uu-\ff} - \epsilon\|W^{-1/2}\qq\|_2.
\end{equation}
Similarly, we can rewrite the VTV penalty function as an optimization over another auxiliary variable, $\zz $. For the channel-by-channel TV, we have
\begin{equation}\label{eq:vtvsdual}
\TV_S(\uu) = \max_\zz \,\, \inner{D \uu,\zz} - \delta_\mathcal{S}(\zz),
\end{equation}
where the definition of the set $\mathcal{S}$ is given by
\begin{equation}\label{eq:setS}
\mathcal{S} = \left\{\zz \in \mathbb{R}^{(\mathcal{M}\times 2)\cdot\mathcal{L}} \,\, : \,\, \mathcal{}\|\left[\zz_m\right]_\ell\|_2 \leq 1 \,\,\, \forall\,\,\, m\in 1,\dotsc,\mathcal{M},\,\,;\ell\in 1,\dotsc,\mathcal{L}   \right\}.
\end{equation}
The quantity $\left[\zz_m \right]_\ell$ refers to image channel $m$ at pixel $\ell$. Similarly, we can write the proposed $\TNV$ penalty function in terms of an optimization over auxiliary variable $\zz$: 
\begin{equation}\label{eq:vtvndual}
\TNV(\uu) = \max_\zz \,\, \inner{D\uu,\zz} - \delta_{N}(\zz),
\end{equation}
where the set $N$ is defined by 
\begin{equation}\label{eq:setN}
N = \left\{\zz \in \mathbb{R}^{(\mathcal{M}\times 2)\cdot\mathcal{L}} \,\, : \,\, \sigma_\mathrm{max}(\left[\zz\right]_\ell) \leq 1 \,\,\, \forall\,\,\, \ell\in 1,\dotsc,\mathcal{L}  \right\}.
\end{equation}
The quantity $[\zz]_\ell$ is an $\mathcal{M}\times2$ matrix, with the same dimensions as the Jacobian derivative at pixel $\ell$, and $\sigma_\mathrm{max}$ is its largest singular value. These transformations, which are detailed in Appendix A, allow us to write our original VTV reconstruction model as a primal-dual, saddle-point problem,
\begin{equation}\label{eq:saddle_qz}
\min_\uu\,\max_{\zz,\qq} \,\, \inner{D\uu,\zz} + \inner{\qq,A\uu-\ff} - \delta_{\mathcal{S}/N}(\zz) - \epsilon\|W^{-1/2}\qq\|_2,
\end{equation}
where the indicator function is either $\delta_{\mathcal{S}}$ for the channel-by-channel $\TV_S$ or $\delta_N$ for the proposed $\TNV$. We can directly apply the CPPD update equations of Algorithm \ref{alg:CPPD}  by making the following assignments.
\begin{equation}
\begin{split}
x &\to \uu,\quad
y \to \begin{pmatrix} \qq \\ \zz \end{pmatrix}, \quad
K \to \begin{pmatrix} A \\ D \end{pmatrix}\\
F(\qq,\zz) &= \delta_{\mathcal{S}/N}(\zz) - \epsilon\|W^{-1/2}\qq\|_2-\inner{\qq,\ff},\quad
G(\uu) = 0\\
\end{split}
\end{equation}
The resulting update equations are given by Algorithm \ref{alg:VTVupdates}.
\begin{algorithm}
\caption{Data-constrained VTV Update Equations}\label{alg:VTVupdates}
\begin{algorithmic}[1]
\State Initialize: $\theta=1$, $\,\, \sigma\tau\left\|\begin{pmatrix}A\\D\end{pmatrix}\right\|_2^2 = 1$, $\,\,(\uu,\uubar, \qq,\zz)\to \bm{0}$ 
\Repeat
\State $\zz^{(k+1)} = \Pi_{\mathcal{S}/N} \left(\zz^{(k)} + \sigma D \uubar^{(k)} \right) $
\State $\qq^{(k+1)} = \prox{\sigma\epsilon F_q}{\qq^{(k)} + \sigma \left(A\uubar^{(k)}-\ff\right)}$
\State $\uu^{(k+1)} = \uu^{(k)} + \tau\left(\Div\zz^{(k+1)} - A^T\qq^{(k+1)} \right)$
\State $\uubar^{(k+1)} = \uu^{(k+1)} + \theta\left(\uu^{(k+1)}-\uu^{(k)} \right)$
\Until  \Comment{stop when convergence criteria met}
\end{algorithmic}
\end{algorithm}
The operator $\Pi_{\mathcal{S}/N}$ is a Euclidean projection onto the set $\mathcal{S}$ or $N$, and the $\Div$ operator is the negative transpose of the discrete derivative operator $D$. Note that the $\qq$-update is written in terms of the proximal map of the function $F_q$, which we define as $F_q = \|W^{(-1/2)}\qq\|_2$. While there does not appear to be a closed-form expression for this proximal map, it can be reduced to a very simple scalar optimization problem. This is detailed in the appendix along with an efficient method for performing the Euclidean projection in the primal variable update. 

\subsection{Simulation studies}

To investigate the impact of the proposed vectorial TV regularization, we performed two numerical simulation studies with the computerized XCAT phantom. In the first experiment we simulate an ideal, photon-counting system with 5 energy windows and directly reconstruct images corresponding to the log-normalized bin data. We will refer to this image basis as the "color" basis because each image channel corresponds directly to the measurements of one energy window. 

In the second experiment, we simulate an ideal, dual-kVp scan and perform what we will refer to as a "hybrid" reconstruction. As we will detail later, we first decompose the 80/140 kVp data into a bone/soft-tissue "material" basis and then we co-reconstruct this synthetic data with the raw 80/140 kVp sinograms. The $\TNV$ penalty couples all four image channels so that the relatively noisy basis-material channels may benefit from the higher SNR color channels. 

\subsubsection{Data generation}

\paragraph{The XCAT shoulder phantom}

\begin{figure}[htbp]
\centering
\newcommand{\pwidth}{0.31\linewidth}
\subcaptionbox{60 keV reference image}{
	\includegraphics[width=\pwidth,trim=0 350 0 500,clip=true]{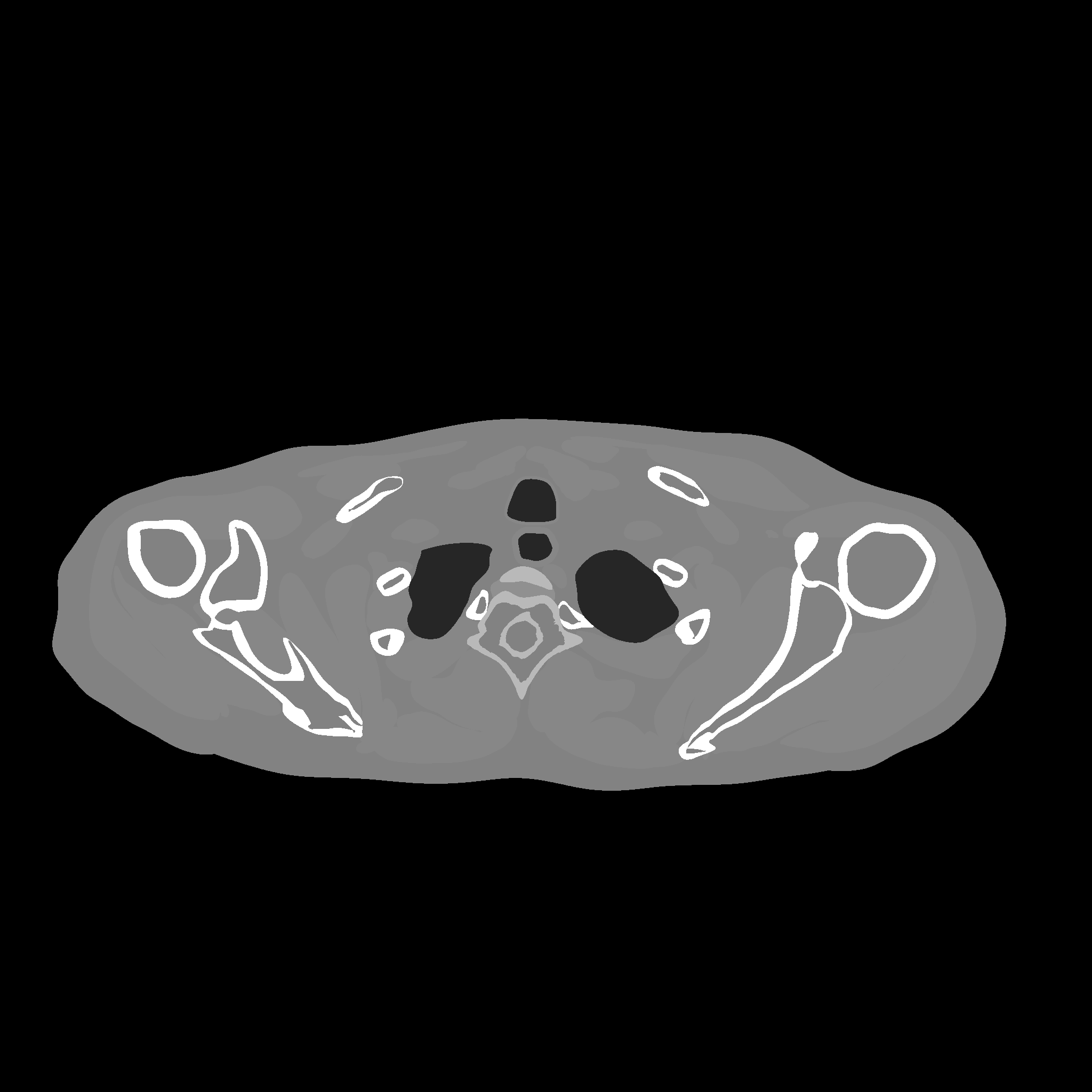}
} \hfill
\subcaptionbox{bone density map}{
	\includegraphics[width=\pwidth,trim=0 350 0 500,clip=true]{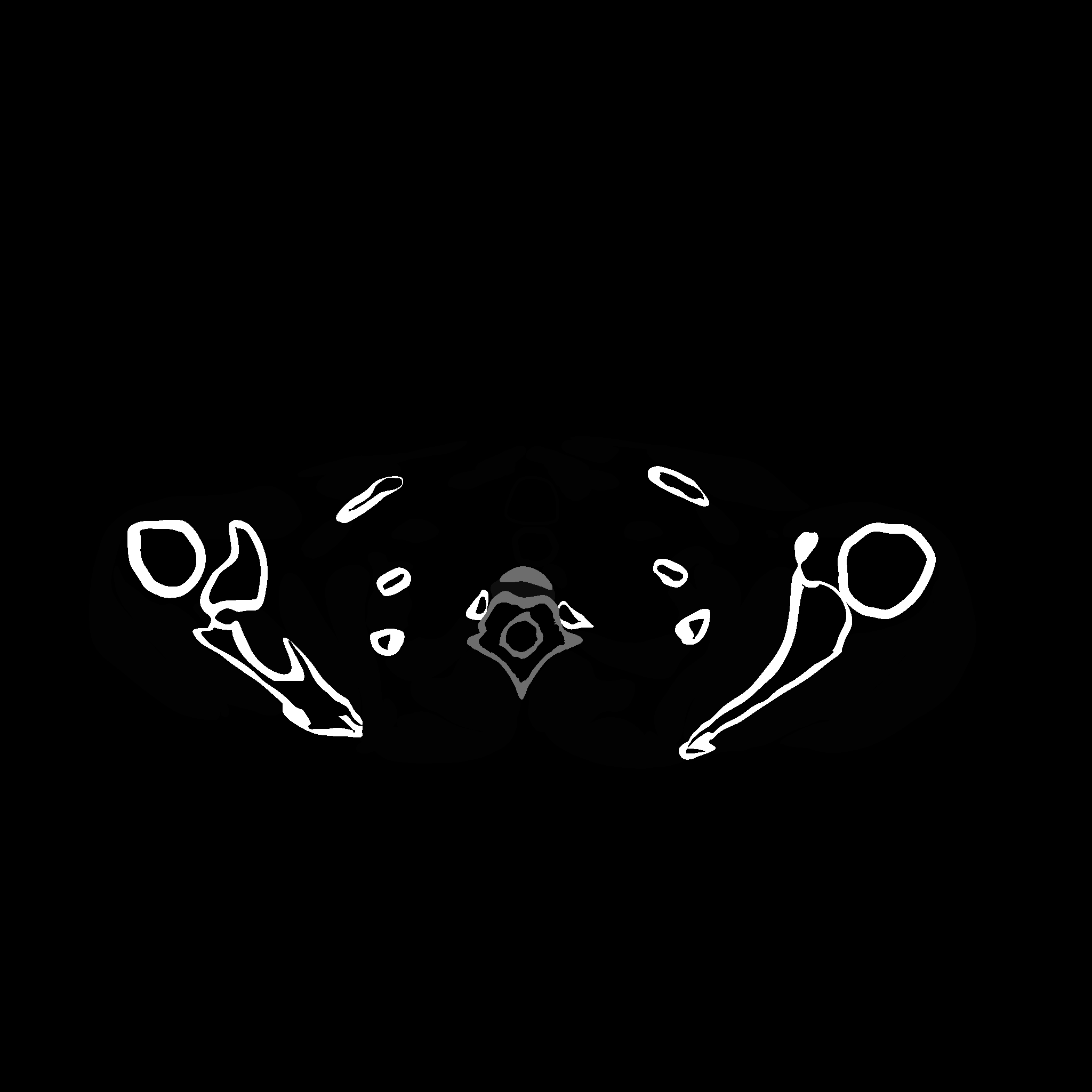}
} \hfill
\subcaptionbox{soft-tissue density map}{
	\includegraphics[width=\pwidth,trim=0 350 0 500,clip=true]{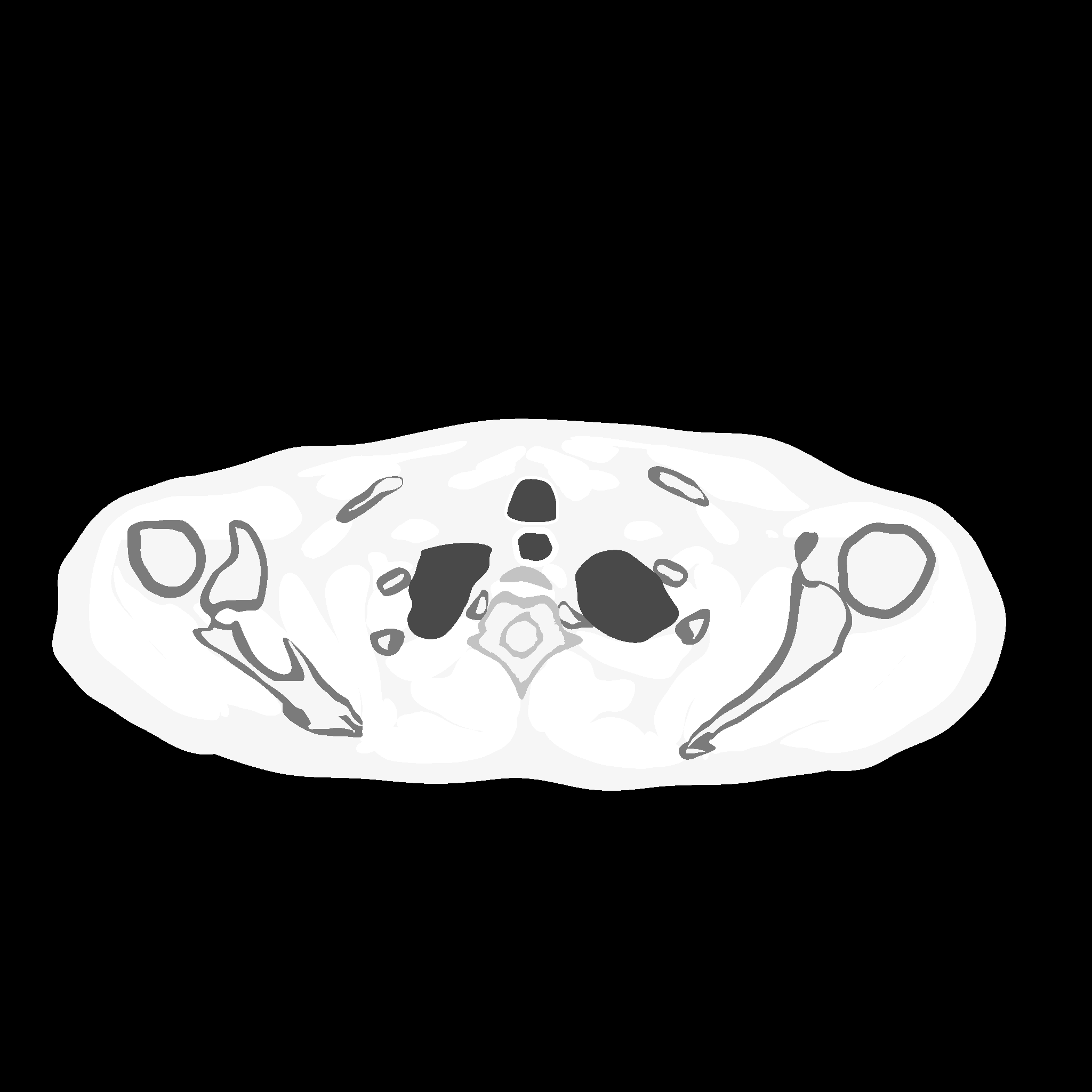}
} \hfill
\caption{The 2D XCAT shoulder phantom used for all simulations studies, depicted at 60 keV (a) for reference. The projection data were generated using bone (b) and soft-tissue (c) density maps with the appropriate mass-attenuation curves.}
\label{fig:XCAT}
\end{figure}

All simulations used the same 2D pixelized shoulder phantom, which was generated from an axial slice of the XCAT phantom and is pictured in Figure \ref{fig:XCAT}. The XCAT software package was used to generated a set of bone and soft-tissue density maps on a $2048\times2048$ pixel grid. These density maps were then used as input to a polyenergetic, distance-driven projector model to generate the simulated raw data that would be inputted to our reconstruction algorithm. 

\paragraph{Ideal photon counting model}

To generate the 5-bin photon-counting data, we used a realistic 120 kVp simulated x-ray tube spectrum with hard thresholds at 40,60,80,100, and 120 keV. We did not model any physical factors in the detector response, so our bin response functions are perfect rect functions. Our simulated spectrum had virtually no emissions below 20 keV, so we can think of these bins as being evenly spaced. Using a distance-driven projector model, polyenergetic projections were computed with 896 detector elements per view and 400 views, and Poisson noise was added. 

\paragraph{Dual kVp hybrid model}
The other configuration we looked at was an 80/140 dual kVp acquisition with the same number of detectors and views as the photon counting simulation. We simulated realistic 80 and 140 kVp x-ray tube spectra to generate two sets of consistent projection data. This type of consistent, dual-kVp data can be acquired on many current scanners by performing two back-to-back scans or can be estimated from a fast kV-switching geometry. Gaussian noise was added to approximate a compound Poisson noise model. From the noisy 80 and 140 kVp projection data, we synthesized bone and soft-tissue basis sinograms using a maximum-likelihood material decomposition. In the "hybrid" study, we will co-reconstruct these 4 channels of data, consisting of our log-normalized, dual kVp data (color basis) and the synthesized, material-basis data. 

\subsubsection{Preprocessing and reconstruction}

Both the 5-channel photon-counting data and the 4-channel dual kVp hybrid data were reconstructed using the reconstruction model outlined in (\ref{eq:constrainedVTV}). For reconstruction, we used a $512\times512$ grid of $1$ mm pixels and a matched projector/backprojector pair based on Joseph's method. 

\paragraph{Computing data-weights}

For the raw-basis projection data, we will use the data-weighting approach described by Bouman \cite{Bouman1996}, which results in a diagonal $W$ matrix, where the diagonal elements equal the pre-logged projection data. This is a quadratic approximation to the log-likelihood. For the synthetic, bone/soft-tissue sinograms, we estimate the variances using the Cramer-Rao lower bound \cite{Roessl2009} and weight by their inverse. This is similar to the approach described by Schirra \cite{Schirra2013} and Sawatzky \cite{Sawatzky2014}, but we ignore the off-diagonal terms in this work for simplicity. We found that having a non-diagonal weighting matrix made it immensely more complicated to derive the algorithm update equations when using this data-constrained form. However, in practice, it is often easier to solve the regularized (unconstrained) problem, and incorporating the off-diagonal covariance terms poses no challenge in this case. We focus on the data-constrained form in this work, simply because it provides a straightforward way to compare the vectorial TV to the channel-by-channel TV.

\paragraph{Tuning the data-constraint parameter}
The only parameter that we will vary in our reconstruction model (\ref{eq:constrainedVTV}) is $\epsilon$, which controls the trade-off between data fidelity and smoothness. In general, smaller values of $\epsilon$ will result in noisier images, while higher values allow the $\VTV$ term to find a smoother solution. At the extreme case of $\epsilon=0$, only images with projections that exactly match the measured data are allowed. There is also some finite value $\epsilon=\epsilon_\mathrm{max}$ for which the algorithm will return an image of all zeros. In order to determine an interesting range for $\epsilon$, we first compute a ground truth image $\utrue$ by performing FBP on noiseless, non-sparse projection data. Then we define a new parameter $\epsilon^*$ which is defined by equation \ref{eq:epsilonstar},
\begin{equation}\label{eq:epsilonstar}
\epsilon^* \equiv \|A\utrue-\ff\|_W
\end{equation}
where $\ff$ are the noisy projection data we wish to reconstruct. In this study we will select values of $\epsilon$ that satisfy $\epsilon = \alpha\epsilon^*$, where $\alpha \in \left(0,1\right)$. We subjectively selected a range of $\alpha$ values that represent a range of solutions from under-smoothed to over-smoothed in order to demonstrate how the TNV compares to the channel-by-channel TV in various regimes. As described earlier, the projection data were re-scaled prior to reconstruction, so that the average noise levels were approximately the same in every spectral channel.

\section{Results}

\begin{figure}[htbp]
\centering
\includegraphics[width=2.5in]{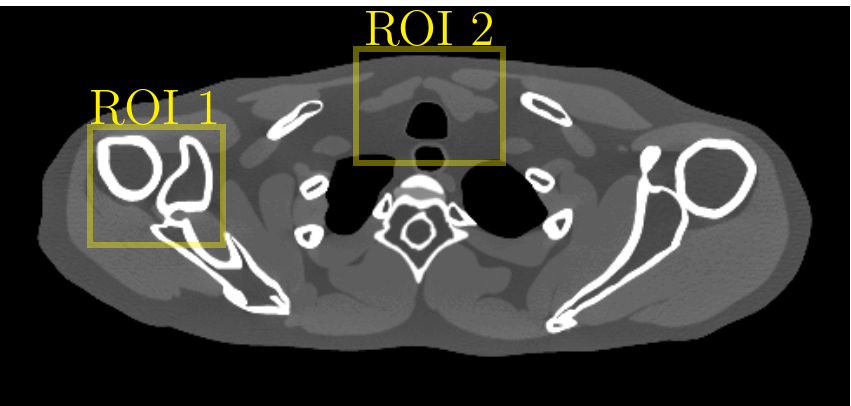}
\caption{XCAT reference image, with ROI's indicated in yellow}
\label{fig:rois}
\end{figure}

\subsection{Photon counting study}

\begin{figure}[htbp]
\centering
\newcommand{\pwidtha}{1in}
\begin{minipage}[t]{0.85\linewidth}
\subcaptionbox{${\TV_S}$\\${\epsilon=0.001\epsilon^*}$\\RMSE$ = 0.0444$}{\includegraphics[width=\pwidtha]{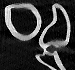}}\hfill
\subcaptionbox{$\TV_S$,\\${\epsilon=0.0013\epsilon^*}$\\RMSE$ = 0.0436$}{\includegraphics[width=\pwidtha]{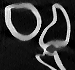}}\hfill
\subcaptionbox{$\TV_S$,\\${\epsilon=0.0016\epsilon^*}$\\RMSE$ = 0.0450$}{\includegraphics[width=\pwidtha]{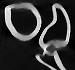}}\hfill
\subcaptionbox{$\TV_S$,\\${\epsilon=0.002\epsilon^*}$\\RMSE$ = 0.0484$}{\includegraphics[width=\pwidtha]{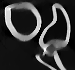}}\hfill
\subcaptionbox{$\TV_S$,\\${\epsilon=0.003\epsilon^*}$\\RMSE$ = 0.0609$}{\includegraphics[width=\pwidtha]{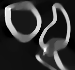}}\hfill

\subcaptionbox{$\TNV$,\\${\epsilon=0.001\epsilon^*}$\\RMSE$ = 0.0390$}{\includegraphics[width=\pwidtha]{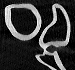}}\hfill
\subcaptionbox{$\TNV$,\\${\epsilon=0.0013\epsilon^*}$\\RMSE$ = 0.0347$}{\includegraphics[width=\pwidtha]{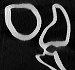}}\hfill
\subcaptionbox{$\TNV$,\\${\epsilon=0.0016\epsilon^*}$\\RMSE$ = 0.0332$}{\includegraphics[width=\pwidtha]{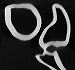}}\hfill
\subcaptionbox{$\TNV$,\\${\epsilon=0.002\epsilon^*}$\\RMSE$ = 0.0327$}{\includegraphics[width=\pwidtha]{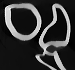}}\hfill
\subcaptionbox{$\TNV$,\\${\epsilon=0.003\epsilon^*}$\\RMSE$ = 0.0456$}{\includegraphics[width=\pwidtha]{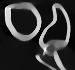}}\hfill
\end{minipage}
\caption{Bin 1 (0-40keV) image, bone ROI/window comparison with channel-by-channel TV (top) and $\TNV$ (bottom). Grayscale window in cm$^{-1}$ $\left[0.30,0.85 \right]$.}
\label{fig:bin1bone}
\end{figure}

\begin{figure}[htbp]
\centering
\newcommand{\pwidtha}{1in}
\begin{minipage}[t]{0.85\linewidth}
\subcaptionbox{$\TV_S$,\\${\epsilon=0.001\epsilon^*}$\\RMSE$ = 0.0219$}{\includegraphics[width=\pwidtha]{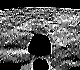}}\hfill
\subcaptionbox{$\TV_S$,\\${\epsilon=0.0013\epsilon^*}$\\RMSE$ = 0.0200$}{\includegraphics[width=\pwidtha]{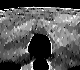}}\hfill
\subcaptionbox{$\TV_S$,\\${\epsilon=0.0016\epsilon^*}$\\RMSE$ = 0.0210$}{\includegraphics[width=\pwidtha]{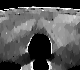}}\hfill
\subcaptionbox{$\TV_S$,\\${\epsilon=0.002\epsilon^*}$\\RMSE$ = 0.0223$}{\includegraphics[width=\pwidtha]{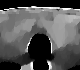}}\hfill
\subcaptionbox{$\TV_S$,\\${\epsilon=0.003\epsilon^*}$\\RMSE$ = 0.0266$}{\includegraphics[width=\pwidtha]{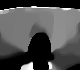}}\hfill

\subcaptionbox{$\TNV$,\\${\epsilon=0.001\epsilon^*}$\\RMSE$ = 0.0216$}{\includegraphics[width=\pwidtha]{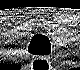}}\hfill
\subcaptionbox{$\TNV$,\\${\epsilon=0.0013\epsilon^*}$\\RMSE$ = 0.0176$}{\includegraphics[width=\pwidtha]{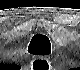}}\hfill
\subcaptionbox{$\TNV$,\\${\epsilon=0.0016\epsilon^*}$\\RMSE$ = 0.0181$}{\includegraphics[width=\pwidtha]{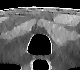}}\hfill
\subcaptionbox{$\TNV$,\\${\epsilon=0.002\epsilon^*}$\\RMSE$ = 0.0189$}{\includegraphics[width=\pwidtha]{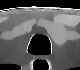}}\hfill
\subcaptionbox{$\TNV$,\\${\epsilon=0.003\epsilon^*}$\\RMSE$ = 0.0175$}{\includegraphics[width=\pwidtha]{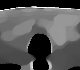}}\hfill
\end{minipage}
\caption{Bin 1 (0-40keV) image,  soft-tissue ROI comparison with channel-by-channel TV (top) and $\TNV$ (bottom). Grayscale window in cm$^{-1}$ $\left[0.30,0.35 \right]$.}
\label{fig:bin1tissue}
\end{figure}

\begin{figure}[htbp]
\centering
\begin{tikzpicture}
\node at (0,0) {\includegraphics[width=3in]{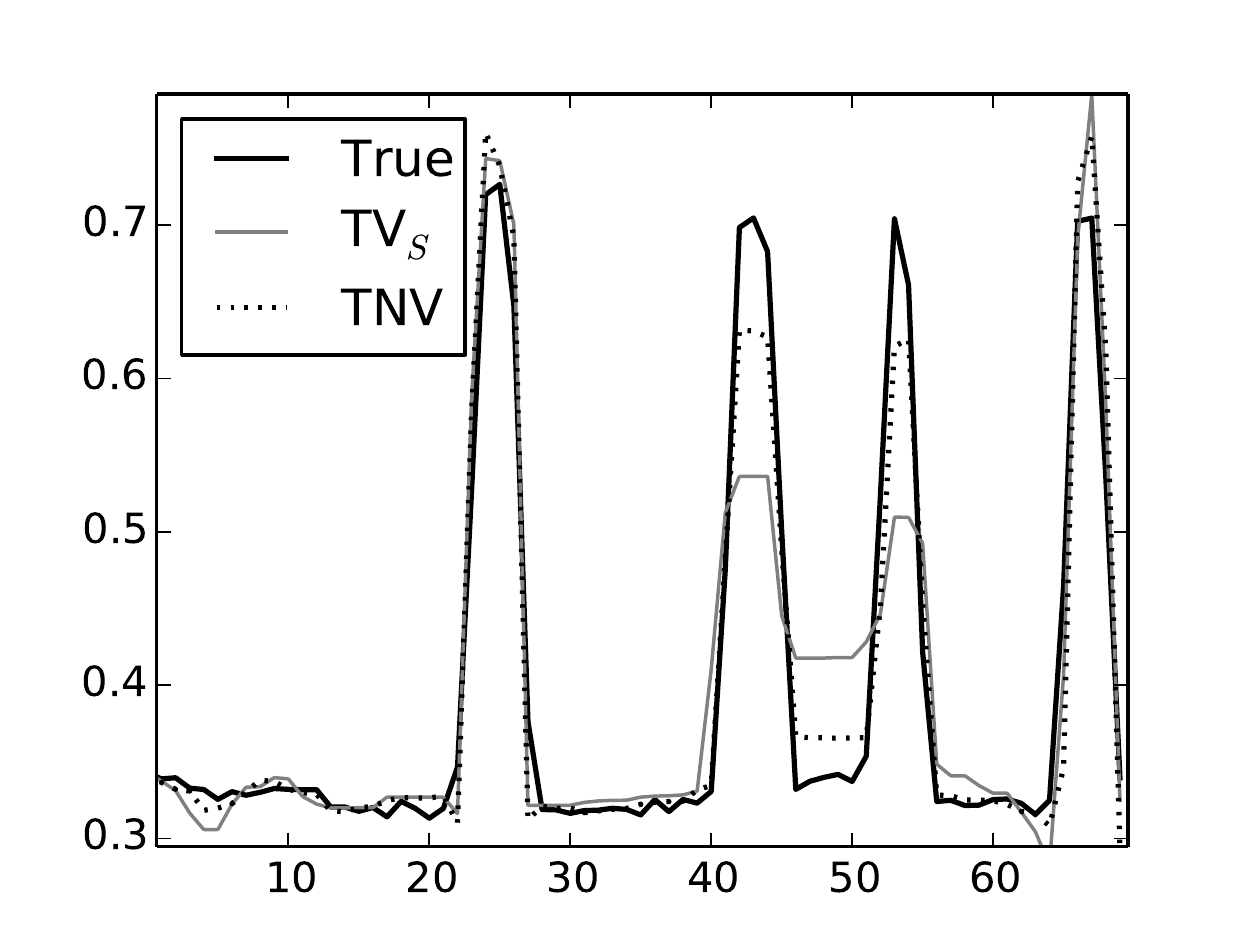}};
\node at (0,-1.3in) {Pixel Index};
\begin{scope}
\node[rotate=90] at (-1.7in,0in) {Attenuation (cm$^{-1}$)};
\end{scope}
\end{tikzpicture}
\caption{This is a vertical profile through the 0-40 keV, bone-ROI image with $\epsilon = 0.0016 \epsilon^*$. The TNV regularized reconstruction shows slightly better preservation of bony structures. The noise levels were estimated from a nearby ROI in a uniform muscle region ($\sigma_{TV_S} = 0.00245$, $\sigma_{\TNV} = 0.00221)$.}
\label{fig:profile_plot}
\end{figure}

Here we present the results of performing a simultaneous reconstruction of the five photon-counting bins after log-normalization, using both the channel-by-channel $\TV_S$ and the proposed $\TNV$. This is an example of performing a joint reconstruction on data in the color basis because we did not perform a material decomposition. In this setup, bin 1 (0-40 keV) was the noisiest, due to significant attenuation below 40 keV. We find that the inter-channel coupling introduced by the total nuclear variation has the greatest impact on the noisiest channels, so we present reconstructed images from this energy bin, focusing on the ROI's indicated in Figure \ref{fig:rois}. The resulting images are depicted in figures \ref{fig:bin1bone} through \ref{fig:bin1tissue}. The image window was manually adjusted for each ROI to highlight the relevant structures but is fixed within a particular figure. In general, we find that the images reconstructed with $\TNV$ regularization suffer from less edge blurring as the smoothing parameter $\epsilon$ is increased. The profile plot in Figure \ref{fig:profile_plot} gives a closer look at how the TNV better preserves bony structures in the lowest energy bin image. We confirmed that these profiles were extracted from images of similar noise levels by measuring the sample variance in a nearby, uniform muscle ROI. This is expected because our data-constrained, reconstruction model allows us to select the noise level directly by tuning the $\epsilon$ parameter. 


\FloatBarrier

\subsection{Dual kVp Hybrid Study}

\begin{figure}[htbp]
\centering
\newcommand{\pwidtha}{1in}
\begin{minipage}[t]{0.85\linewidth}
\subcaptionbox{$\TV_S$,\\${\epsilon=0.06\epsilon^*}$\\RMSE$ = 0.0417$}{\includegraphics[width=\pwidtha]{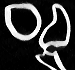}}\hfill
\subcaptionbox{$\TV_S$,\\${\epsilon=0.07\epsilon^*}$\\RMSE$ = 0.0421$}{\includegraphics[width=\pwidtha]{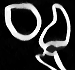}}\hfill
\subcaptionbox{$\TV_S$,\\${\epsilon=0.08\epsilon^*}$\\RMSE$ = 0.0447$}{\includegraphics[width=\pwidtha]{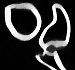}}\hfill
\subcaptionbox{$\TV_S$,\\${\epsilon=0.09\epsilon^*}$\\RMSE$ = 0.0505$}{\includegraphics[width=\pwidtha]{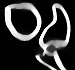}}\hfill
\subcaptionbox{$\TV_S$,\\${\epsilon=0.1\epsilon^*}$\\RMSE$ = 0.0564$}{\includegraphics[width=\pwidtha]{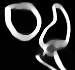}}\hfill

\subcaptionbox{$\TNV$,\\${\epsilon=0.06\epsilon^*}$\\RMSE$ = 0.0410$}{\includegraphics[width=\pwidtha]{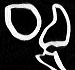}}\hfill
\subcaptionbox{$\TNV$,\\${\epsilon=0.07\epsilon^*}$\\RMSE$ = 0.0398$}{\includegraphics[width=\pwidtha]{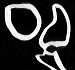}}\hfill
\subcaptionbox{$\TNV$,\\${\epsilon=0.08\epsilon^*}$\\RMSE$ = 0.0392$}{\includegraphics[width=\pwidtha]{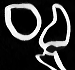}}\hfill
\subcaptionbox{$\TNV$,\\${\epsilon=0.09\epsilon^*}$\\RMSE$ = 0.0392$}{\includegraphics[width=\pwidtha]{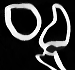}}\hfill
\subcaptionbox{$\TNV$,\\${\epsilon=0.1\epsilon^*}$\\RMSE$ = 0.0405$}{\includegraphics[width=\pwidtha]{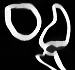}}\hfill
\end{minipage}
\caption{Bone basis image, bone ROI comparison with channel-by-channel TV (top) and $\TNV$ (bottom). Grayscale window in g/mL$^{-1}$ $\left[0.00,0.52 \right]$.}
\label{fig:boneimage}
\end{figure}

\begin{figure}[htbp]
\centering
\newcommand{\pwidtha}{1in}
\begin{minipage}[t]{0.85\linewidth}
\subcaptionbox{$\TV_S$,\\${\epsilon=0.06\epsilon^*}$\\RMSE$ = 0.0445$}{\includegraphics[width=\pwidtha]{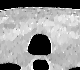}}\hfill
\subcaptionbox{$\TV_S$,\\${\epsilon=0.07\epsilon^*}$\\RMSE$ = 0.0410$}{\includegraphics[width=\pwidtha]{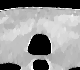}}\hfill
\subcaptionbox{$\TV_S$,\\${\epsilon=0.08\epsilon^*}$\\RMSE$ = 0.0383$}{\includegraphics[width=\pwidtha]{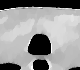}}\hfill
\subcaptionbox{$\TV_S$,\\${\epsilon=0.09\epsilon^*}$\\RMSE$ = 0.0376$}{\includegraphics[width=\pwidtha]{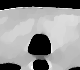}}\hfill
\subcaptionbox{$\TV_S$,\\${\epsilon=0.1\epsilon^*}$\\RMSE$ = 0.0400$}{\includegraphics[width=\pwidtha]{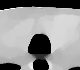}}\hfill

\subcaptionbox{$\TNV$,\\${\epsilon=0.06\epsilon^*}$\\RMSE$ = 0.0431$}{\includegraphics[width=\pwidtha]{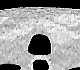}}\hfill
\subcaptionbox{$\TNV$,\\${\epsilon=0.07\epsilon^*}$\\RMSE$ = 0.0396$}{\includegraphics[width=\pwidtha]{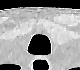}}\hfill
\subcaptionbox{$\TNV$,\\${\epsilon=0.08\epsilon^*}$\\RMSE$ = 0.0391$}{\includegraphics[width=\pwidtha]{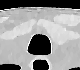}}\hfill
\subcaptionbox{$\TNV$,\\${\epsilon=0.09\epsilon^*}$\\RMSE$ = 0.0386$}{\includegraphics[width=\pwidtha]{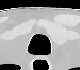}}\hfill
\subcaptionbox{$\TNV$,\\${\epsilon=0.1\epsilon^*}$\\RMSE$ = 0.0372$}{\includegraphics[width=\pwidtha]{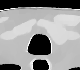}}\hfill
\end{minipage}
\caption{soft-tissue basis image, soft-tissue ROI comparison with channel-by-channel TV (top) and $\TNV$ (bottom). Grayscale window in g/mL$^{-1}$ $\left[0.80,1.07 \right]$}
\label{fig:tissueimage}
\end{figure}

In this study, we co-reconstruct the synthetic soft-tissue/bone image channels with the raw 80/140 kVp data. In general, the material decomposition greatly amplifies noise, so that the soft-tissue and bone images tend to have a much lower CNR than the raw 80 and 140 kVp images. This noise amplification is due to the ill-conditioning of the inversion step in the basis change, which is caused by the relatively poor spectral separation between basis materials. This poor spectral separation also explains the strong negative correlation between the synthesized material channels. We hypothesize that by coupling the high CNR raw image channels and the low CNR synthetic image channels, we may be able to improve noise suppression in the synthetic data. We call this technique of reconstructing the synthetic and raw data simultaneously "hybrid" reconstruction.  We present ROI's from both the bone density image and the soft-tissue density image over a range of different $\epsilon$ values in Figures \ref{fig:boneimage} and \ref{fig:tissueimage}. 
We find that using the TNV allows for a high degree of noise suppression (high values of $\epsilon$) without significantly deteriorating bone or soft-tissue structures. However, when the na\"{i}ve channel-by-channel TV is used, these same $\epsilon$ values eventually lead to significant blurring artifacts. We also point out that even though the soft-tissue, 80 kVp, and 140 kVp images have edges that are not present in the bone density image, the $\TNV$ coupling does not falsely propagate these edges into the bone channel. 

\FloatBarrier

\section{Summary}

We have described a framework for jointly reconstructing multi-channel spectral CT images using a generalized vectorial regularizer. Specifically, we presented a multi-channel generalization of the total variation which couples the different image channels by encouraging their gradient vectors to point in a parallel or anti-parallel direction. Preliminary results suggest that this coupling may allow for greater noise suppression with less blurring of image structures. This regularization strategy can be used to reconstruct raw energy bin data, synthetic basis-material images, or even both simultaneously, such as in our hybrid reconstruction study. This hybrid approach may allow for better noise suppression in the synthetic material images, which typically suffer from elevated noise levels. We believe that the proposed $\TNV$ provides an elegant way to regularize multi-channel images and may be a useful complement to previously published statistical, spectral CT reconstruction methods. 

\bibliography{VTV2}


\clearpage
\appendix
\appendixpage

\section{Deriving the saddle-point problem}

To derive the saddle-point formulation of our optimization problem, we used two fundamental results of convex analysis to "dualize" the data fidelity constraint as well as the VTV term. The transformation of the $\VTV$ term follows straightforwardly from the definition of the so-called "dual-norm," which is defined by
\begin{equation}\label{eq:dualnorm}
\begin{split}
\|x\|' = \sup_z &\,\, \inner{x,z} \,\,\, \mathrm{s.t.} \,\, \|z\|\leq 1.
\end{split}
\end{equation}
Every norm $\|.\|$ has an associated dual-norm $\|.\|'$, that obeys this relationship. The $\ell_2$ norm utilized in the scalar TV is self-dual, while the nuclear norm and spectral norm form a dual pair. 

\paragraph{Dualizing the channel-by-channel TV}
Recall that the channel-by-channel TV can be written as 
\begin{equation}
\TV_S(\uu) = \sum_{m,\ell} \left\|\left[D\uu_m \right]_\ell \right\|_2.
\end{equation}
The definition of the dual-norm allows us to rewrite this as 
\begin{equation}
\begin{split}
\TV_S(\uu) &= \sum_{m,\ell} \sup_\zz \,\, \inner{\left[D\uu_m\right]_\ell,\left[\zz_m\right]_\ell} \,\, \mathrm{s.t.} \,\, \|\left[\zz_m\right]_\ell\|_2 \leq 1\\
&= \sup_\zz \,\, \inner{D\uu,\zz} \,\, \mathrm{s.t.} \,\, \left\|\left[\zz_m\right]_\ell \right\|_2 \leq 1 \,\,\, \forall m \in 1,\ldots,\mathcal{M} \,\,\,; \ell \in 1,\ldots,\mathcal{L}
\end{split}
\end{equation}
which is equivalent to (\ref{eq:vtvsdual}).

\paragraph{Dualizing the proposed TNV}

The proposed TNV is 
\begin{equation}
\TNV(\bm{u}) = \sum_{\ell=1}^{\mathcal{L}} \|\left[D\bm{u}\right]_\ell\|_\star.
\end{equation}
Substituting the definition of the dual-norm, we get
\begin{equation}
\begin{split}
\TNV(\uu) &= \sum_{\ell} \sup_\zz \,\, \inner{\left[D\uu\right]_\ell,\left[\zz\right]_\ell} \,\, \mathrm{s.t.} \,\, \sigma_{\mathrm{max}}\left(\left[\zz\right]_\ell\right) \leq 1\\
&= \sup_\zz \,\, \inner{D\uu,\zz} \,\, \mathrm{s.t.} \,\, \sigma_{\mathrm{max}}\left(\left[\zz\right]_\ell\right) \leq 1 \,\,\, \forall \ell \in 1,\ldots,\mathcal{L},
\end{split}
\end{equation}

which is equivalent to (\ref{eq:vtvndual}). We have used the fact that the spectral norm (maximum singular value) is dual to the nuclear norm. 
\paragraph{Dualizing the data-fidelity term using the Fenchel conjugate}

To dualize the data-fidelity term, we make use of the Fenchel conjugate. For a convex, lower semi-continuous function $f$, the conjugate $f^*$ is defined by 
\begin{equation}\label{eq:fenchel}
f^*(x) = \sup_{x'} \,\, \inner{x,x'} - f(x'),
\end{equation}
where it is also true that $\left(f^*\right)^* = f$. Using this definition, it is easy to show that the following functions form a conjugate pair,
\begin{equation}
\delta_\epsilon(x) \Leftrightarrow \epsilon \|W^{-1/2}x\|_2
\end{equation}
where $\delta_\epsilon(x)$ is the indicator function, defined by (\ref{eq:dataconstraint}). This along with the definition of the Fenchel conjugate leads directly to Eqn. \ref{eq:dataconjugate}.

\section{The proximal map of $\epsilon\sigma\|W^{{-1/2}}\qq\|_2$}
The update equations for Algorithm \ref{alg:VTVupdates} involve evaluating the proximal map of $\epsilon\sigma F_q \equiv \epsilon\sigma\|W^{-1/2}\qq\|_2$, which does not admit a closed form. However, we will now show how it can still be efficiently evaluated using a simple root-finding procedure. First we explicitly write out the proximal mapping as 
\begin{equation}
\prox{\epsilon\sigma F_q}{\qq'} = \argmin_\qq \,\, \|W^{-1/2}\qq\|_2 + \frac{1}{2\epsilon\sigma}\|\qq-\qq'\|_2^2.
\end{equation}
Next, we compare this optimization problem to a slightly easier problem, 
\begin{equation}
\argmin_\qq \,\, \frac{1}{2}\|W^{-1/2}\qq\|_2^2 + \frac{1}{2\lambda\sigma}\|\qq-\qq'\|_2^2,
\end{equation}
and note that for some choice of $\epsilon$ and $\lambda$, these objectives both have the same set of level curves. We can find the solution to this problem by setting the gradient equal to zero, and for a symmetric weighting matrix $W$, it is given by 
\begin{equation}
\qq^* = \left(I + \sigma\lambda W^{-1}\right)^{-1}\qq'.
\end{equation}
If the gradient of this easier problem is equal to that of the original problem, then $\qq^*$ must be an optimizer of the original problem as well. Momentarily, we ignore the non-differentiable point $\qq=\bm{0}$, and set the gradients equal, which yields
\begin{equation}\label{eq:qroot}
\begin{split}
\epsilon &= \lambda \|W^{-1/2}\qq^*\|_2.\\
&= \lambda \|W^{-1/2}\left(I + \sigma \lambda W^{-1}\right)^{-1}\qq'\|_2.
\end{split}
\end{equation}
In this work, we will only consider the case where $W$ is a diagonal matrix, which allows us to simplify this expression.
\begin{equation}\label{eq:lambdaroot}
\epsilon^2 - \lambda ^2\sum_{j=1}^J \frac{W_{jj}}{\left(W_{jj} + \sigma\lambda \right)^2}[\qq']^2_j  = 0.
\end{equation}
The index $j$ is a linear index over every element of $\qq$. Now, we just need to perform a 1D search for a positive value of $\lambda$ that satisfies this equation, which can be done very efficiently using Newton's method or a variety of other algorithms. Once this root, $\lambda^*$, is found, we simply plug it back into (\ref{eq:qroot}):
\begin{equation}
\prox{\epsilon\sigma F_q}{\qq'} = \left\{\begin{array}{lr} \left(I + \sigma\lambda^* W^{-1}\right)^{-1}\qq' & \|W^{1/2}\qq'\|_2 > \sigma\epsilon \\
0 & \|W^{1/2}\qq'\|_2 \leq \sigma\epsilon \end{array}\right..
\end{equation}
Note that because of the non-differentiability of $\|W^{-1/2}\qq\|_2$ at $\qq=0$, there will not always be a solution to equation \ref{eq:lambdaroot}. Specifically, if $\sigma\epsilon > \|W^{1/2}\qq'\|_2$, then there is no viable root. In this case, we need not perform the root-finding procedure, because the optimum must occur at $\bm{q} = \bm{0}$. 

\section{Implementation of the projection operators $\Pi_{\mathcal{S}/N}$}

\paragraph{$\bm{\Pi_\mathcal{S}}$}

Now, we will describe the Euclidean projection onto the set $\mathcal{S}$ defined in (\ref{eq:setS}). Consider projecting $\zz\in\mathbb{R}^{(\mathcal{M}\times 2)\cdot\mathcal{L}}$ onto $\mathcal{S}$. We can define this operation element-wise on each vector $[\zz_m]_\ell \in \mathbb{R}^2$ corresponding to the $m^\text{th}$ spectral channel and the $\ell^\text{th}$ pixel-location. The projection of this element is given by 
\begin{equation}
\Pi_{\mathcal{S}}\left([\zz_m]_\ell \right) = \left\{ \begin{array}{lr} [\zz_m]_\ell  & \|[\zz_m]_\ell\|_2 \leq 0 \\
\frac{[\zz_m]_\ell}{\|[\zz_m]_\ell\|_2} & \text{otherwise}   \end{array}\right..
\end{equation}
For every image channel $m$ and pixel location $\ell$, we simply project the vector $[\zz_m]_\ell$ onto the unit ball. 

\paragraph{$\bm{\Pi_N}$}
The projection onto set $N$ defined by (\ref{eq:setN}) is slightly more complicated. We define this operation element-wise on each $\mathcal{M}\times2$ matrix  $[\zz]_\ell$. This time we need to threshold the eigenvalues of $[\zz]_\ell$. This projection can be succinctly described by
\begin{equation}\label{eq:svd}
\Pi_N\left([\zz]_\ell \right) = U\Sigma_pV^T,
\end{equation}
where $U\Sigma V^T$ is the SVD of $[\zz]_\ell$, and 
\begin{equation}
\Sigma_p = \mathrm{sgn}\left(\Sigma \right)\mathrm{min}\left(|1|,|\Sigma| \right)
\end{equation}
is the thresholded version of $\Sigma$. To form $\Sigma_p$ the singular values on the diagonal of $\Sigma$ are simply thresholded so that their magnitude does not exceed 1. An equivalent projection formula that leads to a much more computationally efficient solution is given by 
\begin{equation}\label{eq:eigenproject}
\Pi_N\left([\zz]_\ell \right) = [\zz]_\ell V \Sigma^\dagger\Sigma_pV^T.
\end{equation}
The quantity $\Sigma^\dagger$ is the pseudo-inverse of $\Sigma$. Since we are only working with 2 spatial dimensions the matrix $V$ will by a $2\times2$ square matrix. Therefore, to compute the projection according to (\ref{eq:eigenproject}) we only need to compute the eigenvalues and eigenvectors of a $2\times 2$ matrix, for which a very simple closed form is available. This can also be done very efficiently for 3D images, where $V$ will be $3\times 3$. Because of this, the update equations that result from the proposed vectorial TV are only trivially more expensive than the channel-by-channel TV, and the projection/backprojection operations are likely to swamp this difference.

\end{document}